\makeatletter \def\namedlabel#1#2{\begingroup
  \def\@currentlabelname{#2}%
  \label{#1}\endgroup
} \makeatother

\newcommand{\papertitle}{Generation and detection of a sub-Poissonian
  atom number distribution\\ in a one-dimensional optical lattice}

\newcommand{\paperkeywords}{Quantum Optics, Fano factor, Nanofiber,
  QND measurement}

\documentclass[twocolumn,amsmath,amssymb, aps, showpacs, showkeys,
floatfix, prl, 10pt,longbibliography]{revtex4-1}

\usepackage[utf8]{inputenc} 
\usepackage[english]{babel} 
\usepackage{csquotes} 

\usepackage{graphicx}
\usepackage{siunitx}%
\usepackage{bm}
\usepackage{xspace}%
\usepackage{xpatch}

\newcommand{\linkcolor}{blue}%
\usepackage{hyperref} 
\hypersetup{colorlinks=true, %
  linkcolor=\linkcolor, %
  citecolor=\linkcolor, %
  filecolor=\linkcolor, %
  urlcolor=\linkcolor, %
  pdftitle=\papertitle, %
  pdfauthor={J.-B. Béguin, E. Bookjans, S. L. Christensen,
    H. L. Sørensen, J. H. Müller, J. Appel, E. S. Polzik}, %
  pdfsubject={Quantum Optics, Quantum noise, Quantum state engineering
    and measurements, Nanofiber}, %
  pdfkeywords=\paperkeywords %
} %

\graphicspath{{images/}}

\usepackage{multibib} 
\newcites{sup}{Supplement Material References}

\newcommand{\nosupplementary}{ %
  \namedlabel{sec:dual-color-homodyne}{\cite{supp}:Dual-color homodyne
    detection}%
  \namedlabel{sec:estimator-variance}{\cite{supp}:Simplified model for
    atom number estimator variance}%
  \namedlabel{sec:recurs-bayes-estim}{\cite{supp}:Recursive Bayesian
    estimation of the atom number}%
  \namedlabel{sec:atom-number}{\cite{supp}:Atom number calibration
    method}
\end{document}\end} %

\renewcommand{\nosupplementary}{} 

\makeatletter%
\def \@labelsection{%
  \@ifundefined{@sectioncntformat}%
  {\@seccntformat}%
  {\@sectioncntformat}{section}%
}%
\def \@labelsubsection{\@labelsection.\thesubsection}%
\def \@labelsubsubsection{\@labelsubsection.\thesubsubsection}%
\xpatchcmd{\@sect@ltx}{\@xsect}{%
  \let\@hskip\hskip%
  \def \hskip { \@hskip 0em plus}%
  \let\@MakeTextUppercase\MakeTextUppercase%
  \def \MakeTextUppercase{}%
  \edef \@currentlabelname{%
    \@hangfrom@section{}{\csname @label#1\endcsname}{#8}%
  } %
  \let\MakeTextUppercase\@MakeTextUppercase%
  \let\hskip\@hskip%
  \@xsect}{}{}
\makeatother

\renewcommand{\Re}{\operatorname{Re}}
\newcommand{\var}{\operatorname{var}}

\newcommand{\diff}[2]{\frac{\mathrm{d}#1}{\mathrm{d}#2}} %
\newcommand{\dd}{\mathrm{d}}%
\newcommand{\e}{\mathrm{e}}

\newcommand{\Phiout}{\Phi_{\text{out}}}
\newcommand{\Phiin}{\Phi_{\text{in}}}
\newcommand{\ket}[1]{\ensuremath{|#1\rangle}\xspace}%
\newcommand{\Nphoton}{N_{\text{ph}}}%
\newcommand{\Natom}{N_{\text{at}}}%
\newcommand{\odatom}{\alpha_{\text{at}}}%
\newcommand{\fref}[2][]{Fig.~\ref{#2}\textcolor{\linkcolor}{#1}} 
\newcommand{\sref}[1]{\nameref{#1}}%
\newcommand{\mean}[1]{\left \langle #1 \right \rangle}

\newcommand{\NBI}{QUANTOP, Niels Bohr Institute, University of
  Copenhagen, Blegdamsvej 17, 2100 Copenhagen, Denmark}

\begin{document}


\newcommand{\correspondingauthors} { \email[Corresponding Authors:
  ]{polzik@nbi.dk} \email{jappel@nbi.dk} \affiliation{\NBI}}

\title{\papertitle}

\author{J.-B. B{\'e}guin}%
\author{E. M. Bookjans}%
\author{S. L. Christensen} %
\author{H. L. S{\o}rensen}%
\author{J. H. M{\"u}ller}%
\author{J. Appel} \correspondingauthors%
\author{E. S. Polzik} \correspondingauthors%

\date{\today}

\begin{abstract}
  We demonstrate preparation and detection of an atom number
  distribution in a one-dimensional atomic lattice with the variance
  $-\SI{14}{\deci\bel}$ below the Poissonian noise level. A mesoscopic
  ensemble containing a few thousand atoms is trapped in the
  evanescent field of a nanofiber. The atom number is measured through
  dual-color homodyne interferometry with a \si{\pico\watt}-power shot
  noise limited probe. Strong coupling of the evanescent probe guided
  by the nanofiber allows for a real-time measurement with a precision
  of $\pm 8$ atoms on an ensemble of some $10^3$ atoms in a
  one-dimensional trap.  The method is very well suited for generating
  collective atomic entangled or spin-squeezed states via a quantum
  non-demolition measurement as well as for tomography of exotic
  atomic states in a one-dimensional lattice.%
  \ifdefined\svnid {%
    \begin{description}%
    \item[SVN] \footnotesize%
      \textcolor{green}{\svnFullRevision*{\svnrev} by
        \svnFullAuthor*{\svnauthor}, 
        Last changed date: \svndate }%
    \end{description}%
  } \fi%
\end{abstract}

\pacs{42.50.Ct, 37.10.Jk, 42.50.Ex}

\keywords{\paperkeywords}
\maketitle

Atoms trapped in an optical lattice are a well-pursued platform for
the realization of a quantum simulator and quantum information
processing devices~\cite{natureBloch}. In addition, mesoscopic
ensembles of periodically well-separated atoms strongly coupled to
light are an excellent arrangement for quantum metrology and sensing
applications using collective atomic state
entanglement~\cite{natphotGiovannetti}.

The recent spectacular progress with cold atoms trapped in the
evanescent field emanating from a tapered optical nanofiber with a
sub-wavelength diameter~\cite{LeKien2004, Vetsch, Kimble,Dawkins2011}
offers a realistic and promising implementation of a one-dimensional
(1D) optical lattice efficiently coupled to a single well-defined
light mode.  Together with the mature technology of interconnecting
optical fibers, atomic ensembles trapped around nanofibers have the
potential to play an integral part in the construction of complex
quantum networks.  Photons propagating in a fiber connect hybrid
quantum systems by interacting with various realizations of quantum
systems, such as solid state systems and atoms, through strong
light-matter coupling at nano-tapered fiber nodes~\cite{hakutaGrating,
  Hafezi}.

An efficient quantum interface between light and collective degrees of
freedom of an atomic ensemble requires a high optical depth and a
measurement sensitivity limited by the shot noise of light and the
projection noise of atoms~\cite{revEugene}. Under these conditions a
quantum non-demolition (QND) measurement of atomic population
differences has been used for the generation of spin
squeezed~\cite{pnasAppel}, entangled states to improve atomic
clocks~\cite{njpAnne,prlVuletic} and
magnetometers~\cite{prlWasilewski,prlMitchell}. In addition, optical
probing of atoms in one-dimensional lattices with sub-Poissonan
precision has been proposed as a valuable measurement tool for
strongly correlated systems~\cite{natureEckert}. The preparation of
ensembles with narrow atom number distribution in atom traps and the
knowledge of its statistics in real-time is also a well-recognized
goal for quantum gate implementations based on collective Rydberg
excitations~\cite{rmpSaffman,Petrosyan2014} or atomic Bragg
mirrors~\cite{njpChang}. For these and other applications it is
desirable to have a probing and preparation method at hand which not
only is minimally-destructive but also widely tunable in
bandwidth. Ideally, it should enable monitoring dynamics on different
timescales and to outrun the influence of any decoherence not caused
by the measurement itself. While impressive atom number resolution has
been reported for atomic ensembles inhomogeneously coupled to an
optical cavity mode~\cite{prlZhang} and for ensembles trapped inside a
low-noise magneto-optical trap~\cite{prlHume}, we demonstrate a fast
single-pass atom number measurement method that is readily adapted to
different measurement and preparation tasks in 1D ensembles.

In this work, we realize the first real-time, minimally-destructive
detection of atoms with sub-Poissonian sensitivity in a 1D nanofiber
lattice trap. Due to the guiding of the probe light by the nanofiber,
the optical depth $\Natom \sigma/A$ achieves maximal values for a
given atom number $\Natom$ as the light beam cross section $A$ becomes
comparable to the atomic cross section. The minimally destructive
measurement is achieved by balancing the phase information obtained
from atoms against the measurement back action, combined with quantum
noise limited sensitivity for both probe photons and atoms. Through a
continuous shot noise limited measurement of the atom induced phase
shift of light, we resolve and prepare an atom number distribution of
the ensemble with a minimum Fano factor $(\Delta \Natom)^2/\Natom$ of
$-\SI{14}{\deci\bel}$.  The reduction of the atom number noise
compared to the Poisson distribution is ultimately limited only by
probe induced stochastic loss of
atoms~\cite{PoissonStochastic,Antoine,PRLSubPoiss}.  The absolute
number of atoms in the lattice trap is calibrated accurately via a
robust experimental method based on optical
pumping~\cite{darkspotKetterle}. Finally, we show that the achieved
light-atom coupling strength and quantum noise limited sensitivity is
suitable for quantum state tomography and will allow for the
generation of many-body entangled states by QND measurements in our
system.

In the experiment Cesium atoms are prepared in a nanofiber trap. Two
counter-propagating red-detuned fields with a wavelength of
$\lambda_\text{red} = \SI{1057}{\nano\meter}$ and a total power of
$P_\text{red}=2\times \SI{1}{\milli\watt}$ together with an
orthogonally linearly polarized blue-detuned running-wave field (with
$\lambda_\text{blue} = \SI{780}{\nano\meter}$,
$P_\text{blue}={\SI{10}{\milli\watt}}$) are sent through an optical
nanofiber and form two one-dimensional optical lattices~\cite{Vetsch}.
With a nominal nanofiber diameter of $d=\SI{500}{\nano\meter}$, the
trapping sites are located $ \SI{200}{\nano\meter}$ above the surface
of the fiber. A magneto-optical trap (MOT) is superimposed on the
fiber; atoms are loaded into the lattice trap after a sub-Doppler
cooling phase, during which they are pumped into the hyperfine ground
state $\ket{3} \equiv (6^2S_{1/2}, F=3)$. Immediately before probing,
the atoms are pumped back to the $\ket{4} \equiv (6^2S_{1/2}, F=4)$
state with external repuming light tuned to the $\ket{3} \to \ket{4'}
\equiv (6^2P_{3/2}, F=4)$ transition.

Our atom number preparation and real-time measurement procedure relies
on the detection of a differential phase shift imprinted by the atoms
on two probe light fields propagating in the fundamental mode of the
fiber. The two probes are detuned symmetrically around the atomic
resonance by $\pm \Omega$, balanced in power and linearly polarized as
the \SI{1057}{\nano\meter} trap field. They are generated from an
acousto-optic modulator (AOM) in the Raman-Nath regime before they are
recombined in a common spatial mode, see \fref[a]{fig:setup}. Due to
the anti-symmetric nature of the atomic
dispersion~(\fref[b]{fig:setup}), the probes have acquired phase
shifts of opposite sign after interacting with the atoms at the fiber
nanotaper.  Measuring the differential phase shift between the two
probes yields a signal proportional to the number of interacting
atoms. At the same time, any common-mode optical path-length and
polarization fluctuation noise is canceled. Furthermore, inhomogeneous
differential Stark shifts imprinted on the atoms by off-resonant
probing are suppressed~\cite{Saffman}.

\begin{figure}
  \includegraphics[width=\columnwidth]{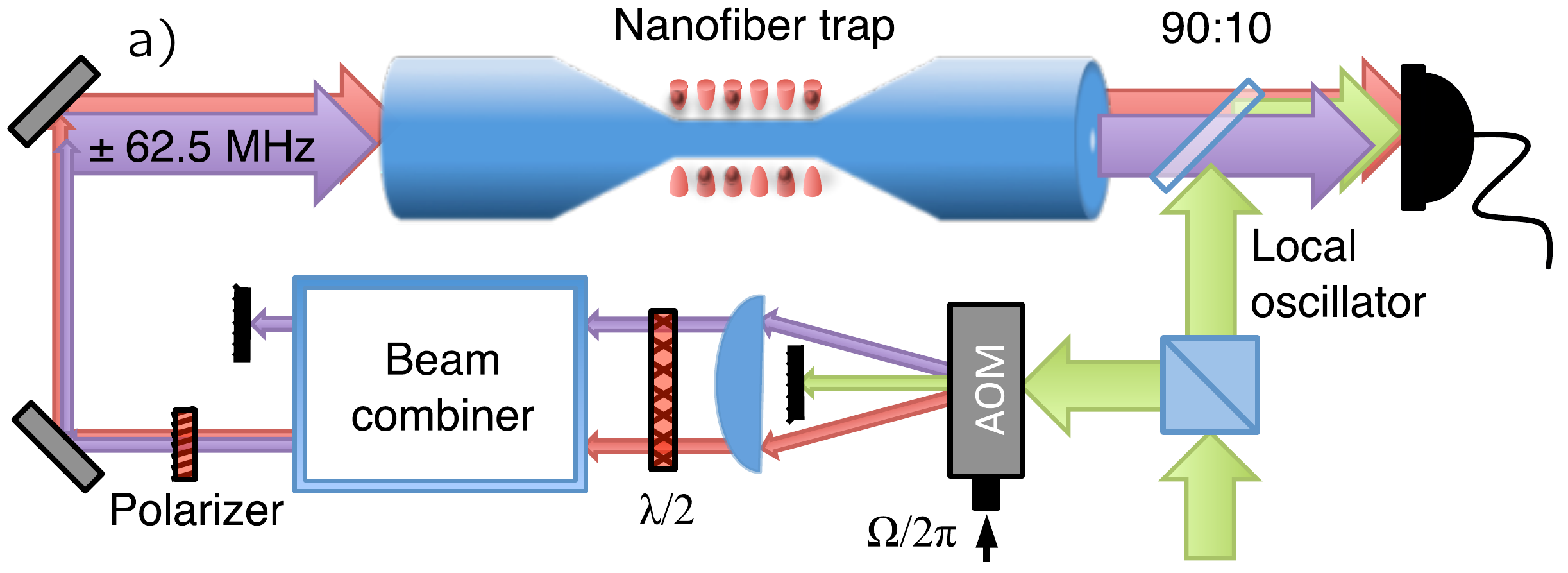}\\
  \includegraphics[width=\columnwidth]{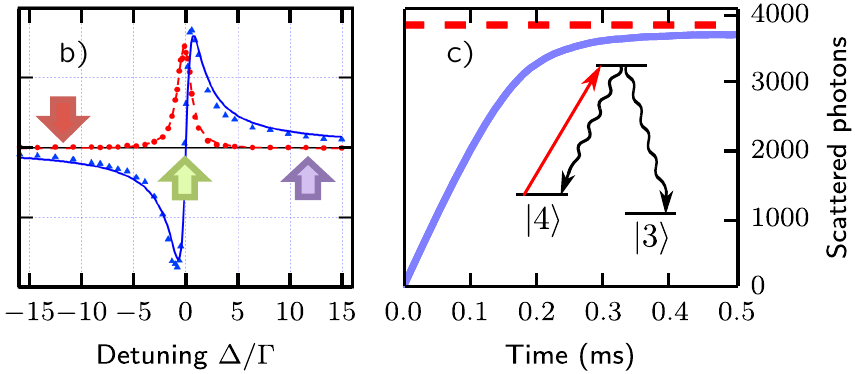}
  \caption{\label{fig:setup}a) Experimental setup.  Two 1D optical
    lattices are formed in the nanofiber trap by two trapping beams
    (not shown).  Two probe beams with fixed frequency difference are
    generated with an AOM and then coupled into the nanofiber where
    they interact with the atomic ensemble. They interfere with a
    strong LO on a 90:10-beam splitter and a homodyne measurement is
    performed.  b) Absorption (red dots) and dispersion signals (blue
    triangles) measured with trapped atoms; arrows indicate probe and
    LO frequencies. c) Atom number calibration. Resonant light pumps
    atoms from~$\ket{4}$ to $\ket{3}$. The total number of trapped
    atoms is determined from the decay branching ratio (see inset) and
    the asymptote (red dashed line) of the cumulative number of
    scattered photons (solid blue line, average of 200 experiments).}
\end{figure}

After passage through the atomic ensemble, the differential phase
shift between the two probes is measured using optical homodyne
interferometry, see~\fref[a]{fig:setup}; the two probes are overlapped
with a strong optical local oscillator (LO) on a $90\text{:}10$ beam
splitter, and the signal is detected with a photo-detector peaked
around the beat-note frequency $\Omega$.  Because of the symmetrical
placement of the probe sidebands with respect to the LO the usual
\SI{3}{\deci\bel} noise penalty for heterodyne detection is
avoided~\cite{josabSUSHI}. The detected beat-note in the photocurrent
is mixed down to baseband electronically and both differential phase
shift and common-mode attenuation of the probe light are extracted
from the signal. All optical fields used for probing are derived from
the same laser source and the optical phase of the LO is stabilized to
the point of highest differential phase sensitivity by a slow servo
loop.

Since in homodyne detection the signal strength and the photon shot
noise contribution from the LO scale identically, technical noise
sources can be overcome in a high detection bandwidth by using a
sufficiently high LO power. The ultimate signal-to-noise ratio (SNR)
with coherent light states is therefore only limited by the intrinsic
quantum noise of the probes. In our experiment, LO photon shot noise
dominates residual electronic noise typically by a factor of 3.5.
This, together with other imperfections, translates into a minimum
phase uncertainty $\delta \varphi$ slightly above the standard quantum
limit expressed as

\begin{align}
  \delta \varphi = \frac{1}{ 2\sqrt{q \Nphoton}}, \quad
  q\equiv\epsilon (1-l) \mathcal{V} \eta. \label{main:eq1}
\end{align}

Here $\Nphoton$ is the total number of probe photons at the atoms
during the measurement probe time and $q$ is an overall quantum
efficiency, dependent on the quantum efficiency of the detector
$\epsilon$, the losses of the probe from the atoms to the detector
$l$, the mode overlap of the probes and LO at the detector
$\mathcal{V}$ and the ratio $\eta$ of the LO shot noise to total
detection noise~\cite{Appel2007}. A value of $q = 0.40 \pm 0.04$ is
achieved in the experimental setup. We have verified the modeled
performance of the detection scheme by a measurement of the
Allan-deviation of interferometer phase in the absence of
fiber-trapped atoms~(for details, see~\sref{sec:dual-color-homodyne}).

By simply blocking one of the probe sidebands, the measurement scheme
turns into standard heterodyne detection, which is used for initial
calibration purposes~\cite{praPino}.  With the remaining sideband
tuned around the $\ket{4} \to \ket{5'} \equiv (6^2P_{3/2}, F'=5)$
transition, the line shape and position of the optical resonance for
trapped atoms is observed (see~\fref[b]{fig:setup}). An energy
absorption measurement with a single probe tuned to the $\ket{4} \to
\ket{4'}$ transition is used to calibrate the number of trapped
atoms. In other nanofiber trap experiments, the number of atoms has
been estimated from the absorbed power of a probe beam fully
saturating the trapped atoms~\cite{Vetsch, Kimble}.  We apply a
similarly robust but faster method, which allows to measure the atom
number in a single run with adequate resolution and good accuracy, by
recording optical pumping transients~\cite{praChen}. Atoms excited to
the $\ket{4'}$ level decay with a fixed branching ratio into the
$\ket{4}$ and $\ket{3}$ ground levels which allows to determine the
number of atoms from the number of absorbed probe photons. From an
average over $178$ consecutive experimental runs, as shown
in~\fref[c]{fig:setup}, we find for the average number of trapped
atoms $\Natom = 1606 \pm 4^{\text{stat}} \pm 160^{\text{sys}}$. The
systematic error for this measurement is dominated by the fractional
uncertainty of the overall quantum efficiency $q$.

To achieve the highest atomic response for the dual-color dispersive
measurement the probes address the atoms in the $\ket{4}$ state
through the excited $\ket{5'}$ state. We detune the probes by $\Omega
= \pm 2\pi \cdot \SI{62.5}{\mega\hertz} \approx \pm 12 \Gamma $ from
the atomic transition, where $\Gamma = 2\pi \cdot
\SI{5.23}{\mega\hertz}$ is the natural line width.  This choice
renders the atomic sample sufficiently transparent to couple all atoms
equally while keeping the influence of neighboring hyperfine levels
small. From the measured phase shift for ensembles with calibrated
atom number we infer an on-resonant optical depth of $\odatom=0.024$
for a single maximally polarized atom on the $\ket{4} \to \ket{5'}$
transition \footnote{We assume here that the dispersively probed
  ensemble is unpolarized.}.  Comparing to earlier results obtained in
a free space optical dipole trap with a related probing
method~\cite{praChristensen}, this represents an improvement of more
than two orders of magnitude in the signal from a single atom.

\begin{figure}
  \includegraphics[keepaspectratio, width=\columnwidth]{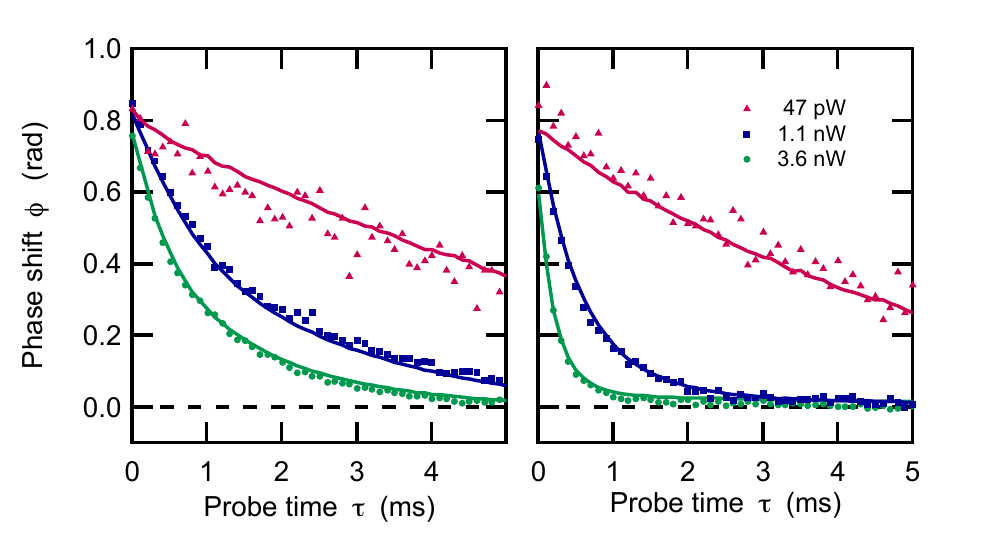}
  \caption{\label{fig:scattering} Real-time high SNR detection of
    atomic phase shift with weak coherent probe light for different
    powers.  (symbols) Real-time phase shift data in a
    \SI{100}{\kilo\hertz} detection bandwidth; (solid lines) Average
    over 200 lattice trap preparations in the same detection
    bandwidth;(left panel) external repumping light on; (right panel)
    external repumping light off; Used probe powers are the same for
    both panels; probing starts \SI{1}{\milli\second} after trap
    loading.}

\end{figure}

\begin{figure}
  \includegraphics[keepaspectratio, width=\columnwidth]{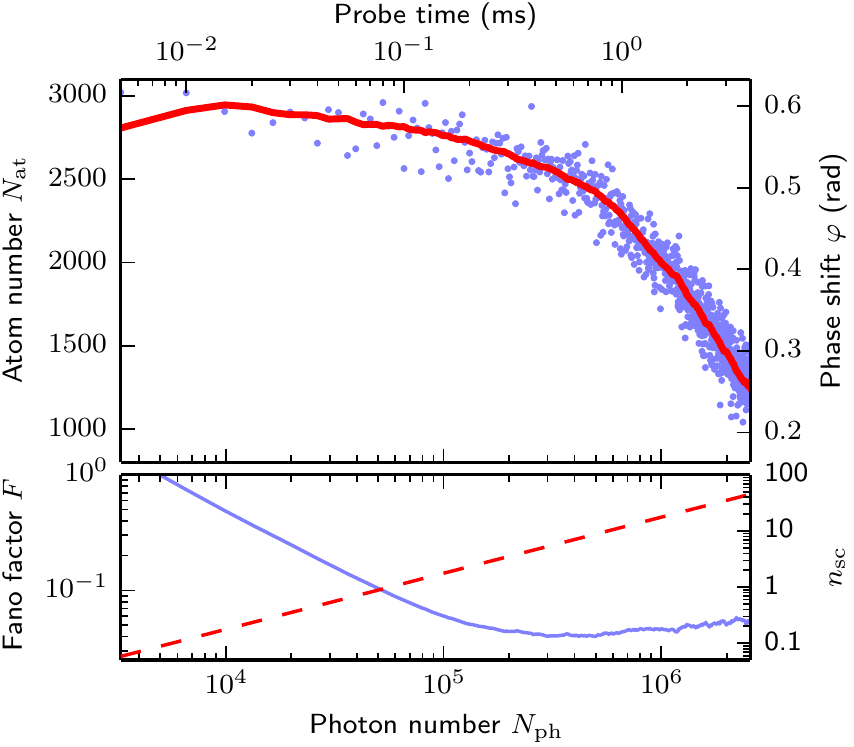}
  \caption{\label{fig:bayesian} Top: Recursive Bayesian estimation of
    the atom number distribution. Blue dots: atom induced optical
    phase shift $\varphi$, as measured in a single experiment with the
    dual-color homodyne technique. Red line: mean of the estimated
    probability distribution for $\Natom$ at every time. Bottom: Solid
    blue line: Fano factor $F=\var(\Natom)/\langle\Natom\rangle$ from
    the same probability distribution. Dashed red line: number of
    scattered photons $n_\text{sc}$ per atom. All data were recorded
    with a probe power of \SI{154}{\pico\watt}, \SI{10}{\milli\second}
    after the MOT cooling phase.}
\end{figure}

To illustrate the wide tunability of strength and bandwidth of the
measurement we show real-time measurements of the atomic phase shift
probing on the $\ket{4} \to \ket{5'}$ transition with and without
external repumping light for varying probe powers
in~\fref{fig:scattering}.  In the shown range the maximum observed
atomic phase shift is independent of the probe power as expected from
the calculated saturation power of $\SI{224}{\nano\watt}$ at the used
probe detuning. The probe-induced signal decay can be made much faster
than the unperturbed trap lifetime without compromising signal
strength by saturating the atoms. The data presented in
\fref{fig:scattering} are taken on a trap with $1/e$ lifetime in the
absence of probing of $\tau_{\text{bg}} = \SI{6.8}{\milli\second}$
\footnote{For the data presented in \fref{fig:bayesian}, intensity
  noise of the trap lasers has been reduced significantly which led to
  a longer trap lifetime of \SI{20}{ms}}. Curiously, we observe that
the probe induced loss rate grows slower than linear with the photon
flux. The average numbers of scattering events $n_\text{heat}$ to
remove an atom from the trap are found to be $n_{\text{heat}} \simeq
380$ for $P = \SI{3.6}{\nano\watt}$, while $n_{\text{heat}} \simeq
190$ for $P = \SI{1.1}{\nano\watt}$ and $n_{\text{heat}} \simeq 56$
only for $P = \SI{0.15}{\nano\watt}$. Plain recoil heating in
individual trap sites of calculated depth $10^3$ $E_{\text{recoil}}$
predicts constant $n_{\text{heat}} \simeq 500$~\cite{Wolf2000} and
clearly cannot explain the data. This peculiar behavior is not
understood at present and is subject to further studies.  In the
absence of repumping light, probe-induced hyperfine pumping into the
atomic state $\ket{3}$ is observed on average after $n_{\text{hf}} =
67$ spontaneous emission events, in good agreement with the calculated
value at the used probe detuning.

We now apply the calibrated dispersive minimally-destructive probing
method to prepare atom number distributions in the optical lattice
with sub-Poissonian fluctuations. The Fano factor quantifies the
reduction in the atom number fluctuations as compared to a Poisson
distribution as $F = (\Delta \Natom)^2/\mean{\Natom}$. A Fano factor
below unity is sometimes referred to as number squeezing.  In the
experiment atoms are probed \SI{10}{\milli\second} after the
sub-Doppler cooling to avoid transit signals from untrapped atoms from
the initial MOT reservoir.  Atoms are probed on the $\ket{4}
\rightarrow \ket{5'}$ transition and external repumping light on the
$\ket{3}\rightarrow \ket{4'}$ transition is used to counteract
hyperfine pumping.  In \fref{fig:bayesian} we show a typical record of
the measured real-time phase shift from a single realization where
data points are averaged over $\SI{5}{\micro\second}$. The noisy data
is seen to follow a smoothly decaying curve with time or equivalently
the probe photon number.  We apply a recursive Bayesian estimation
procedure to track the atom number distribution at a given invested
probe photon number $\Nphoton$ from all phase measurement data up to
that time (see~\sref{sec:recurs-bayes-estim}): We describe our
knowledge of the atom number by an initially uniform probability
distribution.  With every sample of acquired, shotnoise-contaminated
data, we update this distribution using Bayes rule of inference. In
every step, we evolve our estimator distribution to account for
stochastic loss of atoms due to background gas collisions and due to
heating by the probe light. The lower panel in \fref{fig:bayesian}
displays the Fano factor for the atom number estimator. We find a
minimum Fano factor of $-\SI{14}{\deci\bel}$ from the knowledge
acquired by \num{5E5} probe photons which led to a loss of only
\SI{14}{\percent} of the initial atoms. This demonstrates that we can
prepare ensembles with arbitrary atom numbers between $1000-2500$ with
Fano factors well below $-\SI{10}{\deci\bel}$.

For the preparation of the total trapped atom number, a strong
measurement can be applied, since the only back-action mechanism
changing the variable of interest is atom loss due to recoil heating.
Tomographic characterization of collective atomic hyperfine coherence
by measuring atomic population differences instead does not allow for
the use of repump light~\cite{njpChristensen}.  Preparation of
spin-squeezed ensembles by a QND measurement is even more stringent
and limits the number of allowed spontaneous emission events below
unity~\cite{pnasAppel,prlVuletic}.

We use a simplified model for the variance of the atom number
estimator, inspired by~\cite{prlZhang}, in order to evaluate the
potential of the measurement scheme for the different tasks. Assuming
that all atom loss is caused by probe light, the variance of the
estimator writes as
\begin{align}
  (\Delta \Natom)^2 =\left(\frac{1}{(\Delta \Natom^\text{i})^2} + q
    \alpha_{\text{at}} n_{\text{sc}}\right)^{-1} + \Natom
  \frac{n_{\text{sc}}}{n_{\text{loss}}}. \label{main:eq2}
\end{align}
Here, $\alpha_{\text{at}}$ denotes the single-atom optical depth on
the probe transition, $n_{\text{sc}}$ is the number of probe photons
scattered into free space for a single atom, $(\Delta
\Natom^\text{i})^2$ is the initial variance of the atomic ensemble
before probing, and $n_{\text{loss}}$ the critical number of
scattering events, i.e. $n_{\text{heat}} = 56$ for atom number
preparation as in \fref{fig:bayesian}, $(n_{\text{hf}}^{-1} +
n_{\text{heat}}^{-1})^{-1}$ for state tomography and $n_{\text{loss}}
\lesssim 1$ for conditional spin-squeezing. Any initial information
about the atom number distribution is encoded in the prior variance
$(\Delta \Natom^\text{i})^2$.  The first term describes the gain of
knowledge from the phase shift measurement while the second term
reflects the noise from stochastic atom loss.

The measurement strength characterized by $n_{\text{sc}}$, can now be
optimized for all three tasks (see~\sref{sec:estimator-variance}). For
atom number preparation we find $n_{\text{sc}} = 2.4$ leading to a
predicted minimum Fano factor of $\SI{-11}{\deci\bel}$ for the
parameters of \fref{fig:bayesian}. The simple model is somewhat
pessimistic but still in reasonable agreement with the observed
values.

In quantum state tomography where reduction of the variance has to be
balanced with the noise introduced by heating and hyperfine pumping,
the minimum predicted Fano factor is $\SI{-8}{\deci\bel}$.  This
should be compared with $\SI{-3}{\deci\bel}$ required to observe
negative Wigner function ensemble distributions and hence allows
characterization of non-classical spin states for ensembles containing
$2500$ atoms.  Extrapolating to measurement-based preparation of
spin-squeezed collective atomic states, where we need to take the
damping of hyperfine coherence due to photon scattering into account,
we find that metrologically relevant squeezing up to
$\SI{-4.2}{\deci\bel}$ in our system is achievable if all other
decoherence channels are negligible.

In conclusion, we have demonstrated an efficient interface between
fiber-guided light modes and atomic ensembles trapped in a 1D optical
lattice. The nanofiber trap geometry offers two obvious routes for
future improvements. The single atom coupling strength can be
moderately increased by pulling atoms closer to the fiber surface, but
also substantially increased by embedding the trap into an optical
resonator using integrated fiber Bragg gratings~\cite{Wuttke2012,
  hakutaGrating}. Alternatively, the ensemble size can be increased by
simply using longer fiber sections without compromising the single
atom coupling.  At least one order of magnitude larger ensembles are
realistic with current state of the art nanofiber production
technology.

\begin{acknowledgments}
  We gratefully acknowledge Prof. A.~Rauschenbeutel and his group for
  advice and access to their fiber pulling rig and thank
  Prof. M.~F.~Andersen for fruitful and enlightening discussions. This
  work has been supported by the ERC grant INTERFACE, the US ARO grant
  No. W911NF-11-0235 and the EU grant SIQS.
\end{acknowledgments}

\bibliographystyle{apsrev4-1} 
\bibliography{articlebib}
\label{LastBibItem}

\nosupplementary

\newpage 

\setcounter{page}{1} 

\appendix

\setcounter{equation}{0} 
\renewcommand{\theequation}{S\arabic{equation}} 
\renewcommand{\theHequation}{\theequation} 

\setcounter{figure}{0} 
\renewcommand{\thefigure}{S\arabic{figure}} 
\renewcommand{\theHfigure}{\thefigure} 

\section*{Supplemental Material} \renewcommand{\appendixname}{SM}

\section{Dual-color homodyne detection}\label{sec:dual-color-homodyne}

We consider an idealized balanced homodyne setup with a strong local
oscillator field $E_\text{LO}$ and a dual-color probe field
$E_\text{signal}$ impinging on a $50:50$ beam splitter. Both sidebands
experience an phase shift by $\varphi$ with opposite
sign:\begin{align}
  E_\text{LO} & = E_\text{L} e^{i \omega t} \\
  E_\text{signal} & = E_{1} e^{i (\omega+\Omega)t + i \varphi} %
  + E_{2} e^{i (\omega-\Omega)t - i \varphi}\label{eq:3}
\end{align}

From the fields at the two beam splitter output ports $E_\pm =
(E_\text{LO} \pm E_\text{signal})/\sqrt{2}$ we obtain the intensities:
\begin{align}
  I_\pm & = |E_\pm|^2 = \frac{1}{2} \left| E_\text{LO}\right|^2 +
  \frac{1}{2} \left| E_\text{signal}\right|^2 \pm
  \Re{\left(E_\text{LO}^* E_\text{signal}\right)}\label{eq:4}
\end{align}

The homodyne detector signal is obtained by subtracting the
photocurrents in both arms and therefore it is proportional to
\begin{align}
  \Delta I & = I_+ - I_- = 2 \Re{\left(E_\text{LO}^*
      E_\text{signal}\right)}.\label{eq:5}
\end{align}

Inserting (\ref{eq:3}) we obtain
\begin{align}
  \Delta I & = 2 \Re \Bigl[ & & E_\text{L}^* (E_1 + E_2) \cos(\Omega t
  + \varphi) \nonumber \\ & & + & i E_\text{L}^*(E_1 - E_2)
  \sin(\Omega t + \varphi) \quad \Bigr].\label{eq:6}
\end{align}

We choose the local oscillator phase to be real
$E_\text{L}=\sqrt{I_\text{LO}}$ and balance the sideband powers, so
that $E_1=E_2=\sqrt{I_\text{s}/2}$ and obtain a beat signal at
frequency $\Omega/2 \pi$ phase shifted by the optical sideband phase
shifts $\varphi$:
\begin{align}
  \Delta I(t) = 2 \sqrt{2 I_\text{s} I_\text{LO}} \cos(\Omega t +
  \varphi).\label{eq:7}
\end{align}

For phase detection, we integrate this signal over a time $\tau= 2 \pi
m /\Omega$ corresponding to $m$ oscillation periods while we
demodulate with $\sin(\Omega t)$.  Designating $\kappa$ as the
conversion factor from optical intensities to photo-electron-flux the
number of detected and demodulated photo-electrons is:
\begin{align}
  \Delta n_\tau^\text{sin} & = \int_0^\tau \kappa \Delta I(t) \sin(\Omega t) \, \dd t \\
  & = \kappa 2 \sqrt{2 I_\text{s} I_\text{LO}}  \int_0^\tau \cos(\Omega t+\varphi) \sin(\Omega t) \, \dd t \\
  & = -\kappa \tau \sqrt{2 I_\text{s} I_\text{LO}} \sin
  \varphi.\label{eq:8}
\end{align}

Shot noise in the photocurrent limits the obtainable phase resolution
to
\begin{align}
  \delta \phi = \frac{\delta \Delta n_\tau^\text{sin}}{ \left|
      \frac{\dd \Delta n_\tau^\text{sin}}{\dd \varphi}
    \right|}, \label{eq:9}
\end{align}
where $\delta \Delta n_\tau^\text{sin}$ denotes shot noise
fluctuations of $ \Delta n_\tau^\text{sin}$. The fluctuations of the
$\cos$- and $\sin$ demodulated components are equal in power and
independent, and the powers add up to that of the vacuum fluctuations:
$(\delta \Delta n_\tau^\text{sin})^2 +(\delta \Delta
n_\tau^\text{cos})^2= (\delta \Delta n_\tau)^2$.  Since $I_\text{s}\ll
I_\text{LO}$, in the balanced detection $\delta \Delta n_\tau=\delta
n_\tau = \sqrt{n_\tau}$ where $n_\tau= \kappa \tau I_\text{LO}$ is the
total number of detected photons during the time $\tau$. We therefore
have
\begin{align}
  \delta \Delta n_\tau^\text{sin} = \delta \Delta n_\tau^\text{cos} &
  = \frac{1}{\sqrt{2}} \sqrt{n_\tau} = \sqrt{\kappa \tau
    I_\text{LO}/2}.\label{eq:10}
\end{align}

\begin{figure}
  \includegraphics[width=\columnwidth]{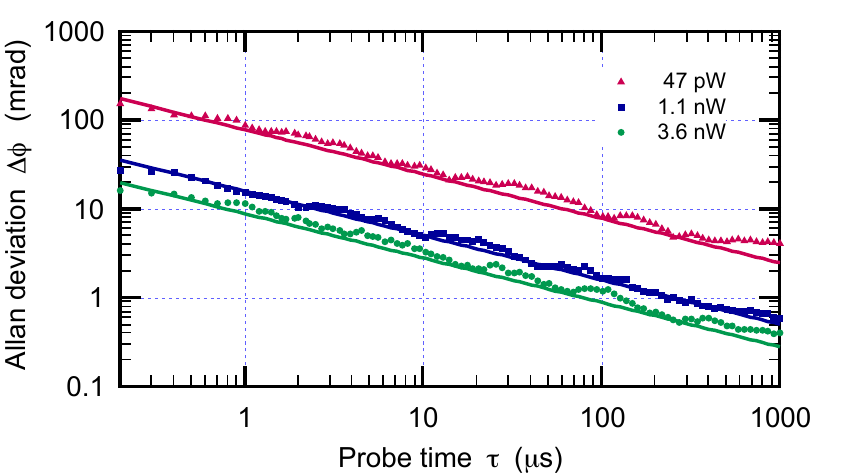}
  \caption{\label{fig:noise_graph} Log-log scale of the measured Allan
    deviation of the phase as a function of the probe time window
    $\tau$ for different probe powers at the location of the lattice
    trap. (Points) Experimental data (Solid lines) Expected
    theoretical quantum noise for the the overall quantum efficiency
    of detection $q = 0.40$ and probe power.}
\end{figure}

From equations~(\ref{eq:9}) and~(\ref{eq:10}) we now obtain for the
shot noise limited phase resolution in the dual-color homodyne
setting:
\begin{align}
  \delta \varphi & = \frac{\sqrt{\kappa \tau
      I_\text{LO}/2}}{\kappa\tau\sqrt{2 I_\text{s} I_\text{LO}}} =
  \frac{1}{2\sqrt{\kappa \tau I_\text{s}}} =
  \frac{1}{2\sqrt{n_\text{s}}}.\label{eq:11}
\end{align}
Here $n_\text{s} = \kappa \tau I_\text{s}$ is the total number of
detected signal photons. We note that this result is identical with
the limiting resolution in homodyne detection of the phase of a
single-frequency coherent state with a mean photon number of
$n_\text{s}$.  In the main text we express the phase resolution in
terms of the photon number sent through the atomic ensemble leading to
eq.~\eqref{main:eq1}.  To test the predicted minimal light shot noise
limited phase resolution $\delta \varphi$, we measure the Allan phase
deviation, $\Delta \phi$, in absence of atoms for different probe
light powers (see \fref{fig:noise_graph}). We find excellent agreement
between measured and expected Allan deviation with the independently
measured detection efficiency $q$.

\section{Recursive Bayesian estimation of the atom number}
\label{sec:recurs-bayes-estim}
The method of recursive Bayesian~\citesup{bookbayesian} estimation
provides a framework to estimate a probability distribution for the
number of atoms during the process of probing based on the cumulative
knowledge obtained by weak measurements. It is based on dividing the
probing into discrete time intervals; in each interval the optical
phase induced by the atoms is sampled. At the same time, the atom
number in the probed level changes due to processes such as collisions
with the background gas, probe-induced heating, optical pumping, etc.

In each time step $l$ of duration $\Delta t$, we probe the ensemble
with $\Delta N$ probe photons; the atom number evolution is modeled as
a Markov process, so that each atom has a fixed probability $P$ to
``disappear'', i.e. the probability to have $N_l$ atoms in the trap in
the $i$-th step given that the atom number was $N_{l-1}$ in the
previous step is governed by a binomial distribution:
\begin{align}
  p(N_l | N_{l-1}) &= \binom{N_{l-1}}{N_{l}} (1-P)^{N_l}
  P^{N_{i-l}-N_{l}},\label{eq:12}
\end{align}
so that on average the atom number evolves as
\begin{align}
  \langle N_l \rangle & = N_0 e^{-l P} = N_0 e^{- \frac{t}{\Delta t}
    P}. \label{eq:13}
\end{align}

Also in each step, we obtain a new measurement of the optical phase
shift $\varphi$ proportional to the current atom number $N_l$ with an
added random Gaussian shot noise contribution $\varepsilon$ with $\var
\varepsilon= (\delta \varphi)^2$:
\begin{align}
  \varphi_l & = k N_l + \varepsilon,\\
  \intertext{so that $\varphi_l$ is normally distributed:}
  p(\varphi_l|N_l) & = \frac{1}{\sqrt{2 \pi \var{\varepsilon}}}
  e^{-\frac{1}{2} \frac{(\varphi_l - k N_l)^2}{\var
      \varepsilon}}.\label{eq:14}
\end{align}

Starting with an initial estimation of the atom probability $p(N_0)$,
in each step we use the result of the new phase measurement
$\varphi_l$ to update our estimation:
\begin{align}
  p(N_i | \varphi_{1 \ldots l}) & = \alpha \, p(\varphi_l|N_l) \,%
  p(N_{l} | \varphi_{1 \ldots l-1}),\label{eq:15}
\end{align}
where $\alpha$ normalizes $\sum_{N=0}^\infty p(N | \varphi_{1 \ldots
  l})=1$ and
\begin{align}
  p(N_{l} | \varphi_{1 \ldots l-1}) & = \sum_{N_{l-1}=0}^\infty
  p(N_l|N_{l-1}) p(N_{l-1} | \varphi_{1 \ldots l-1}).\label{eq:16}
\end{align}

\begin{figure}
  \centering
  \includegraphics[width=\columnwidth]{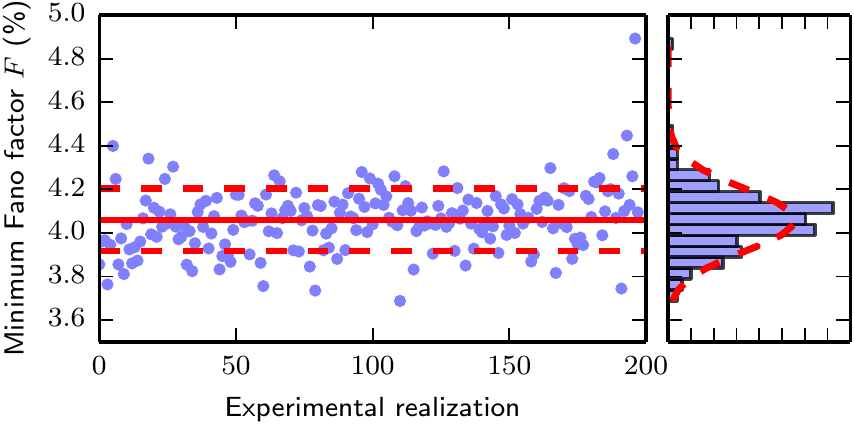}
  \caption{Left:~minimum Fano factor $F$ obtained in 200 successive
    experiments. Right:~histogram.}
  \label{fig:FanoStatistics}
\end{figure}

For~\fref{fig:bayesian}, we assume a uniform probability $p(N_0)$ for
$N_i<4400$ and calibrate $P$ from fitting eq.~(\ref{eq:13}) to the
mean atom number decay recorded in 200 separate experiments. In
Fig.~\ref{fig:FanoStatistics} we present the statistics of the
achieved Fano factors obtained by Bayesian estimation of the atom
number in those 200~experiments.

\section{Atom number calibration method} \label{sec:atom-number}

\begin{figure}
  \includegraphics[keepaspectratio, width=\columnwidth]{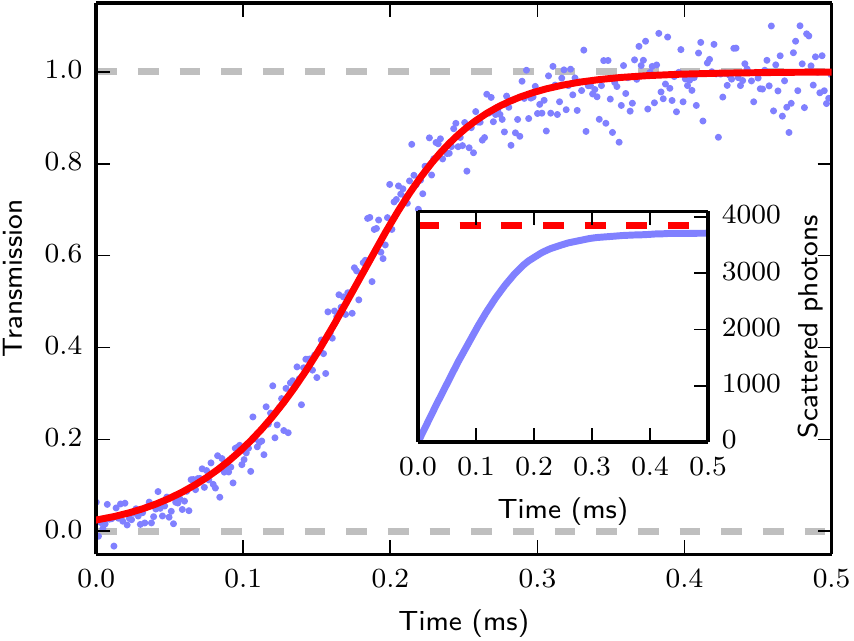}
  \caption{\label{fig:atomnumber} Calibration measurement of the
    absolute atom number in the lattice trap.  An atom number of
    $\Natom = 1606 \pm 4^{\text{stat}} \pm 160^{\text{sys}}$ atoms is
    determined through optical pumping into a dark state via the
    integration of the total number of probe photons required to
    bleach the ensemble. The probe light is resonant with the atomic
    $\ket{4} \rightarrow \ket{4'}$ transition and the fiber-guided
    probe power at the atoms is \SI{5.0}{\pico\watt}. The transmission
    is deduced from an heterodyne measurement of the two quadratures
    of the probe field. Signals are recorded \SI{10}{\milli\second}
    after the sub-Doppler cooling phase. The shown data is an average
    over 178 runs. The model fit estimates an associated on-resonance
    optical depth per atom of $\SI{1.64}{\percent}$ on the closed
    transition $(\ket{4,m_F=4} \rightarrow \ket{5',m_F=5})$.}
\end{figure}

To accurately determine the number of trapped atoms a single sideband
resonant with the $\ket{4} \to \ket{4'}$ transition is used. In brief,
the idea behind the measurement is as follows. An excited atom in any
of the $\ket{4'}$ states can decay either to the $\ket{3}$ and
$\ket{4}$ manifold. The branching ratio for this decay is independent
of the Zeeman $m_{F'}$ sublevel.  Because of the large ground state
hyperfine splitting Cs atoms in the $\ket{3}$ state are far-detuned
from the probe laser frequency and therefore do interact with the
probe light only very weakly, i.e., atoms in this state do \emph{not}
absorb any photons from the probe to an excellent approximation. The
number of trapped atoms in $\ket{4}$ will decrease with time during
probing and the probe transmission will increase accordingly. To find
the number of atoms we need to determine the number of scattering
events it takes for the probe to reach full transmission. To model
this, we consider a simple three level scheme and derive a
differential equation for the optical depth, $d$, as this will
directly give the transmission using the Lambert-Beer law. The number
of atoms in the $\ket{4}$ state $N_4$ changes due to scattering
events. We have with $\Phiin (\Phiout)$ as the input (output) photon
flux respectively,
\begin{align}
  \diff{N_4}{t} = -\frac{1}{k} \Bigl(\Phiin (t) - \Phiout(t)
  \Bigr).\label{eq:17}
\end{align}
On average, it takes $k=2.4$ scattering events to pump an atom from
$\ket{4}$ to $\ket{3}$ as can be easily derived from the partial decay
rates ($7/12\Gamma$, $5/12\Gamma$) of the $\ket{4'}$ states. The
output photon flux follows Lambert-Beer's law
\begin{align}
  \Phiout(t) & = \Phiin(t) \exp (-d(t)), \intertext{where} d(t) &
  \equiv \Natom (t) \odatom~\label{eq:18}
\end{align}
is the optical depth, and $\odatom$ denotes the per-atom optical depth
on the used transition. Combining eqs.~(\ref{eq:17}) to (\ref{eq:18}),
we obtain the desired differential equation for $d(t)$:
\begin{align}
  \diff{}{t}d(t) = - \frac{\odatom}{k} \Phiin(t) \left[1 - \exp (-
    d(t)) \right],\label{eq:19}
\end{align}
Assuming a constant input flux, the solution is
\begin{align}
  d(t) = \ln \left[ 1+\left( \e^{d(t=0)}-1 \right)\exp \left( -
      \odatom \Phiin t/k \right) \right],\label{eq:20}
\end{align}
from which we find for the sample transmission $T(t)=\e^{-d(t)}$ the
expression for the fit model in~\fref{fig:atomnumber}.
\begin{align}
  \label{eq:transmission}
  T(\Natom, \odatom, t) = \frac{1}{ 1+ \left[ \exp \left( \odatom
        \Natom \right)-1 \right]\exp \left( -\odatom \Phiin t/k \right
    )};
\end{align}
allowing us to deduce both $\Natom$ and $\odatom$.

Since the method is used to calibrate the atom number, possible
systematic effects need to be addressed. Probe laser detuning,
polarization, population redistribution among Zeeman sub-levels and
inhomogeneous broadening of the probe transition by trap light do not
influence the number of spontaneous emission cycles needed to bleach
the sample and hence do not change the estimated atom number. The
possible dark state in the Zeeman manifold of the $\ket{4}$ level,
where population could be trapped, is avoided by a suitably oriented
magnetic field. Modifications to the excited state decay rate caused
by the close proximity to the dielectric surface of the fiber are
small (below the percent level) and not expected to change the
branching ratio to first order. Other systematic effects like
radiation trapping, which has compromised the accuracy of the method
in previous implementations with cold atoms, as well as collective
back-scattering are largely suppressed by the extreme 1D geometry and
the incommensurate lattice and probe wavelengths. In practice, the
largest systematic error by far stems from the calibration uncertainty
of the overall quantum efficiency $q$ and hence the input photon flux
to the atomic sample, which amounts to $10\%$ in our experiment. We
note that this calibration error is not subject to fluctuations
between consecutive experimental runs.

Several of the above mentioned parameters and effects systematically
lower the \emph{speed} of the optical pumping transient and hence bias
the value for the fitted single atom coupling strength
$\odatom$. Similarly, atom number variations from run to run conspire
to soften the sharp transition from almost complete absorption to full
transmission when raw signals are averaged over many realizations. We
observe a significant difference between coupling strength extracted
from a fit to an averaged trace, as shown in~\fref{fig:atomnumber},
and the average coupling strength from fits to individual traces. The
atom number estimate, instead, is robust against interchange of
fitting and averaging operations. Comparing the average of inferred
coupling strength values from individual fits to the coupling strength
from dispersive phase shift measurements we find a remaining relative
deviation of $25\%$.

\section{Simplified model for atom number estimator
  variance}\label{sec:estimator-variance}

To arrive at the formula for the variance of the atom number estimator
in eq.~\eqref{main:eq2} we start by relating the phase resolution of
the homodyne method eq.~\eqref{main:eq1} to the atom number resolution
given a number of photons scattered into free space
$n_{\text{sc}}$. Coherent probe light passing through an ensemble of
2-level atoms with on-resonance optical depth $d_0 =
\alpha_{\text{at}} \cdot \Natom$ acquires a phase shift $\varphi$
\begin{align}
  \label{eq:2LevelPhase}
  \varphi = \frac{d_0}{2} \frac{\tilde{\Delta}}{1 +
    \tilde{\Delta}^2}\simeq \frac{d_0}{2} \tilde{\Delta}^{-1} ,
\end{align}
with $\tilde{\Delta}$ the detuning in units of half the natural
linewidth $\Gamma/2$ and the approximation valid for the
experimentally relevant case of $\tilde{\Delta} \gg 1$. In the same
situation the probe photon flux $\dot{N}_{\text{ph}}$ (equivalently
photon number $N_{\text{ph}}$ in a stationary situation) is reduced as
\begin{align}
  \label{eq:2LevelAbsorption}
  \frac{\Delta N_{\text{ph}}}{N_{\text{ph}}} = d_0 \frac{1}{1 +
    \tilde{\Delta}^2}\simeq d_0 \tilde{\Delta}^{-2} .
\end{align}

Combining the two equations we get for an ``ensemble'' containing a
single atom:
\begin{align}
  \label{eq:2LevelSingle}
  \varphi_1^2 = \frac{\alpha_{\text{at}}}{4}
  \frac{n_{\text{sc}}}{N_{\text{ph}}} ,
\end{align}
where $\varphi_1$ denotes the phase shift due to a single atom.

The variance of the estimator for differential phase shift in a
measurement using the dual-color homodyne method follows from
eq.~\eqref{main:eq1}. Together with the phase shift of an ensemble
containing $\Natom$ atoms $\varphi = \Natom \cdot \varphi_1$ this
translates into the variance of the atom number estimator as
\begin{align}
  \label{eq:EstimatorVariance}
  (\Delta\varphi)^2 & = \frac{1}{4 q N_{\text{ph}}} & \Rightarrow &
  (\Delta \Natom)^2 = \frac{1}{4 q N_{\text{ph}} \varphi_{1}^2}
  \intertext{and upon using eq.~\eqref{eq:2LevelSingle}} (\Delta
  \Natom)^2 & = \frac{1}{q \alpha_\text{{at}} n_{\text{sc}}}.
\end{align}

As expected for a dispersive measurement the variance does \emph{not}
depend on the atom number.  Low variance is achieved with high quantum
efficiency $q$, high single atom coupling strength
$\alpha_{\text{at}}$ and a high average scattered photon number. The
estimator variance is seen to diverge when no measurement is
performed, i.e.  when $n_{\text{sc}} \to 0$. From a Bayesian
estimation perspective this corresponds to an uninformative
prior. When doing a measurement, the variable of interest ($\Natom$)
can change due to the back-action of the meter (probe light and vacuum
modes) on the system (atomic ensemble).  We incorporate Poissonian
atom loss due to probe light scattering into free space and prior
information about the atom number distribution, leading to
\begin{align}
  (\Delta \Natom)^2 =\left(\frac{1}{(\Delta \Natom^\text{i})^2} + q
    \alpha_{\text{at}} n_{\text{sc}}\right)^{-1} + \Natom
  \frac{n_{\text{sc}}}{n_{\text{loss}}}. \label{suppl:eq2}
\end{align}
as stated in eq.~\eqref{main:eq2} of the main text.

A useful measurement reduces the variance of the estimator
substantially with respect to the prior variance. Discarding the prior
information for now, we optimize the measurement strength for a single
step estimation of the atom number. Note that a single step estimation
procedure performs worse than than the quasi-continuous Bayesian
filtering method presented above, since the full variance of atom loss
contributes despite the continuous monitoring.  This explains the
pessimistic predictions of the simplified estimator variance model.
Working out the appropriate derivative, the minimum variance is
achieved for a number $n_{\text{sc}}$ of
\begin{align}
  n_{\text{sc}} & = \left(\frac{n_{\text{loss}}}{\Natom q
      \alpha_{\text{at}}}\right)^{1/2} , \intertext{leading to minimum
    variance and minimum Fano factor as}
  (\Delta \Natom)^2_{\text{min}}  & = \left(\frac{4 \Natom}{n_{\text{loss}} q \alpha_{\text{at}}}\right)^{1/2}\\
  F_{\text{min}} & = \left(\frac{4 }{\Natom n_{\text{loss}} q
      \alpha_{\text{at}}}\right)^{1/2}.
  \label{OptimalF}
\end{align}
The numbers cited in the main text for atom number preparation
(tomography) are calculated using $\Natom = 2500$ ($\Natom = 1250$)
and $n_{\text{loss}} = 56$ ($n_{\text{loss}} = 30$) in
eq.~\eqref{OptimalF}.  We assume here that tomographic
characterization of the collective hyperfine coherence is performed
with only $50\%$ of the population residing in the coupled level.

For the extrapolation to measurement induced spin-squeezing the
allowed measurement strength is so weak that the prior information
becomes significant.  In addition, for metrologically relevant
squeezing the loss of hyperfine coherence (reduced Ramsey fringe
contrast) needs to be taken into account properly, while the loss of
atoms during the measurement is negligible for the weak measurement.
We assume a coherent spin state of an ensemble with $\Natom$ 2-level
atoms in the equatorial plane of the collective Bloch sphere as the
initial condition. A weak measurement of the population of the upper
level $\ket{4}$ is performed. Given this, the prior variance of the
upper level population is $(\Delta \Natom^\text{i})^2 = \Natom/4$.
Scattering of probe photons reduces the contrast of fringes in a
Ramsey experiment and hence the discriminator slope for a parameter
measurement by a factor $\exp{(-n_{\text{sc}}/2)}$, since only half of
the atoms are probed.  We take eq.~\eqref{suppl:eq2}, neglect the atom
loss term, normalize with the variance of the prior and correct for
the loss of fringe contrast to get for the achievable squeezing $\xi$
\begin{align}
  \xi \equiv \frac{(\Delta \Natom)^2}{\Natom/4} \e^{n_{\text{sc}}} & =
  \frac{1}{1 + q \alpha_{\text{at}} \Natom n_{\text{sc}}/4}
  \e^{n_{\text{sc}}}
\end{align}
or, in terms of the optical depth of the ensemble
\begin{align}
  \xi = \frac{1}{1 + q d_0 n_{\text{sc}}/4} \e^{n_{\text{sc}}}.
\end{align}

In the limit of high optical depth $d_0$ the squeezing is optimized
for $n_{\text{sc}} = 1$, equivalent to a reduction of Ramsey fringe
contrast to $60\%$. To achieve any useful squeezing at all the product
of optical depth and quantum efficiency needs to satisfy $q d_0 >
4$. For the numerical example of squeezing performance given in the
main text, we use an ensemble size of $\Natom = 2500$ and assume atoms
probed on the cycling transition leading to $\xi =
\SI{-4.2}{\deci\bel}$. This scenario is relevant for magnetic field
sensing using the $\ket{4,4}$ to $\ket{3,3}$ hyperfine coherence.

\bibliographystylesup{apsrev4-1} 
\bibliographysup{articlebib}

\end{document}